\documentclass[fleqn,usenatbib]{mnras}

\usepackage{newtxtext,newtxmath}

\usepackage[T1]{fontenc}
\usepackage{ae,aecompl}

\usepackage{graphicx}	
\usepackage{amsmath}	
\usepackage{amssymb}	

\usepackage{hyperref}

\usepackage[usenames, dvipsnames]{color}
\usepackage{soul}

\usepackage[export]{adjustbox}[2011/08/13]

\title[NGC-3175 Satellites]{Faint Satellite Population of the NGC-3175 Group - a Local Group Analogue}

\author[R. Kondapally et al.]{
Rohit Kondapally$^{1,2}$\thanks{E-mail: rohitk@roe.ac.uk},
George A. Russell$^{2}$, Christopher J. Conselice$^{2}$ \newauthor{ and Samantha J. Penny$^{3}$}
\\
$^{1}$SUPA, Institute for Astronomy, University of Edinburgh, Royal Observatory, Edinburgh, EH9 3HJ, UK\\
$^{2}$Department of Physics and Astronomy, University of Nottingham, Nottingham, NG7 2RD, UK\\
$^{3}$ Institute of Cosmology and Gravitation, University of Portsmouth, Dennis Sciama Building, Burnaby Road, Portsmouth PO1 3FX, UK
}

\date{Accepted XXX. Received YYY; in original form ZZZ}

\pubyear{2018}

\begin{document}
\label{firstpage}
\pagerange{\pageref{firstpage}--\pageref{lastpage}}
\maketitle

\begin{abstract}
In this paper we identify and study the properties of low mass dwarf satellites of a nearby Local Group analogue - the NGC-3175 galaxy group with the goal of investigating the nature of the lowest mass galaxies and the `Missing Satellites' problem.  Deep imaging of nearby groups such as NGC-3175 are one of the only ways to probe these low mass galaxies which are important for problems in cosmology,
dark matter and galaxy formation.  We discover 553 candidate dwarf galaxies in the group, the vast majority of which have never been studied before.  We obtained R and B band imaging, with the ESO 2.2m, around the central $\sim$500kpc region of NGC-3175, allowing us to detect galaxies down to $\sim$23 mag (M\textsubscript{B}$\sim$-7.7 mag) in the B band. In the absence of spectroscopic information, dwarf members and likely background galaxies are separated using colour, morphology and surface brightness criteria. We compare the observed size, surface brightness and mass scaling relations to literature data. The luminosity function with a faint end slope of $\alpha$ = -1.31, is steeper than that observed in the Local Group. In comparison with simulations, we find that our observations are between a pure $\Lambda$CDM model and one involving baryonic effects, removing the apparent problem of finding too few satellites as seen around the Milky Way.
\end{abstract}



\section{Introduction}\label{chap:introduction}
	The $\Lambda$ cold dark matter ($\Lambda$CDM) model has enjoyed considerable success in explaining the observed properties of structure formation and growth. This model matches with the observed fluctuations in the cosmic microwave background power spectrum \citep{hinshaw2013nine,planck2014cosmology}, the growth of large scale structure, quasar absorption lines and the Lyman $\alpha$ forest \citep{springel2006large}. Because of this, one main focus over the past two decades has been to constrain or modify $\Lambda$CDM by searching for deviations from observations at small scales, such as within galaxy groups.

	In a hierarchical scenario, galaxies assemble via mergers of smaller dark matter (DM) haloes where the merging process is not entirely smooth, i.e. substructures (haloes) are not always destroyed. Numerical simulations of pure dark matter particles indeed predict thousands of DM haloes orbiting the Milky Way (MW) and the Local Group (LG) compared to the few tens and hundreds, respectively, of observed satellite galaxies \citep{kauffmann1993formation,moore1999dark,klypin1999missing,springel2008Aquarius} - this is known as the ``missing satellites problem''. In particular, it was found by \citet{klypin1999missing} that the discrepancy in the abundance of satellites between the hierarchical models and observations occurs below v\textsubscript{circ} $\sim$ 50kms\textsuperscript{-1}, corresponding to the faint end of the luminosity function. The study of dwarf galaxy population is therefore an important probe of the hierarchical $\Lambda$CDM model at small scales. 

	Since it was first shown that substructure would likely survive in galactic haloes, there has been considerable effort towards trying to reconcile $\Lambda$CDM with observations. Theories have been focused on producing realistic MW halo simulations by invoking various baryonic physics including winds and, stellar and supernova feedback effects all of which, suppress star formation in low mass galaxies (e.g., \citealt{scanna2009discsdata,wadepuhl2011satellite,scanna2012aquila}). In addition, feedback from active galactic nuclei (AGN) has also been suggested to suppress star formation, even in the most massive dwarf galaxies \citep{dashyan2018agn}. 
    
    Although the most recent generation of cosmological simulations (e.g., EAGLE project: \citealt{schaye2015eagle,crain2015eagle-issues} and APOSTLE: \citealt{fattahi2016apostle}) benefit from the increase in computational power, higher resolution and the incorporation of feedback effects, there remains a need to implement the uncertain subgrid physics. For example, \cite{scanna2012aquila} compare the formation of a MW analogue using 13 different cosmological simulation codes in a $\Lambda$CDM structure formation scenario. Each code is run with the same initial conditions, but the outputs exhibit large variations in stellar mass, size, morphology and gas content.
    
    Observationally, the census of Local Group satellites has been largely complete at the bright end, and thus the focus over the past decade has shifted to detecting ever fainter satellites (M\textsubscript{B} $>$ -9). Systematic searches of the Local Group have revealed such faint satellites around both M31 \citep{2007ApJ...671.1591I,2007ApJ...659L..21Z,2008ApJ...676L..17I} and a population of ultra faint and ultra diffuse galaxies around the Milky Way \citep{belakurov2006bootes,zucker2006canes,koposov2015beasts,kim2016pegasus3,torrealba2016giant,torrealba2016aquarius2}, with now $>$ 100 known satellites of the LG.
    
    Whilst other possible solutions to the missing satellites problem have been proposed, using for example, Warm dark matter cosmologies, which can suppress the number of low mass haloes that form galaxies, it is possible that the satellites of the Milky Way and the Local Group may not be representative of their mass scale. Additionally, observations have shown that the Magellanic Clouds may have a satellite system of their own (e.g., \cite{koposov2015beasts, 2018MNRAS.475.5085T}) which has introduced debate on the membership of some newly discovered satellites to the LMC or the Milky Way. Proper motions from Gaia data release 2 find some of the newly discovered dwarf galaxies to be inconsistent with being associated to the Magellanic Clouds \citep{2018arXiv180501448K}. This ongoing debate in the literature suggests a need to study abundance and properties of satellites around analogues, beyond the Local Group.
    
	In the nearby universe, studies of satellite properties of hosts with different masses to the Milky Way have already begun  (e.g., NGC 253: \citealt{sand2014discovery}; NGC 3109: \citealt{sand2015antlia}; NGC 6503: \citealt{koda2015discovery}; NGC 2403: \citealt{carlin2016first}; M101: \citealt{merritt2014m101,bennet2017m101dw}; M81: {\citealt{2013AJ....146..126C}}; Centaurus A: {\citealt{2016ApJ...823...19C}}). A large survey is also being led to study the satellite population around 100 Milky Way analogues - the SAGA survey \citep{geha2017saga_arXiv}. Initial results from 8 such analogue hosts find the existence of the missing satellites problem. However, this survey has a magnitude limit of M\textsubscript{r} $\sim$ -12.3, whereas many satellites fainter than this limit have been found around the Milky Way, where the differences between models and observations is the most acute.
    
    In this study, we focus on the NGC-3175 group - a nearby LG analogue. This group was chosen based on both the similarity of K-band luminosities of the two large galaxies to the Local Group, and the group's local environment. The group consists of two large spiral galaxies, NGC-3137 (M\textsubscript{K} = -22.2) and NGC-3175 (M\textsubscript{K} = -22.9) \citep{skrutskie2006_2mass}. In comparison, the M31 has M\textsubscript{K} = -23.4, which is computed using $m_{K}$ = 1.1 mag \citep{2003AJ....125..525J} and a distance of 784kpc \citep{1998ApJ...503L.131S}. The MW has M\textsubscript{K} = -24.0 \citep{1996ApJ...473..687M, 2001ApJ...556..181D}. In addition, the group also contains other low mass spiral galaxies, comparable to the LG. Our aim is to identify the satellite population, study their properties and carry out an initial comparison with the predictions from $\Lambda$CDM.

	This paper is organised as follows: Section 2 provides an overview of the imaging data from the ESO WFI telescope. Section~\ref{sec:3} describes the SExtractor configuration used for object detection and star-galaxy separation to generate a sample of potential dwarf candidates. Section~\ref{sec:4} describes the implementation of GALFIT, using GALAPAGOS-2 to fit surface brightness profiles of the sample, and examines the model fitting quality through comparisons with simulated objects. We also determine here the criteria for defining group membership of the dwarf galaxy sample. We analyse the properties of the dwarf candidates, in comparison with observations from literature data in Section~\ref{sec:5}. The cosmological implications of our findings are explored in Section~\ref{sec:6} by the way of comparisons to various $\Lambda$CDM simulations of the Local Group. We provide our conclusions in Section~\ref{sec:7}.
    
    In this paper, we use a $\Lambda$CDM cosmology with $\Omega_{m} = 0.3$, $\Omega_{\Lambda} = 0.7$ and $H_{0}=70$ h kms\textsuperscript{-1} Mpc\textsuperscript{-1}. We use a distance of 14Mpc to NGC-3175 \citep{sorce2014dist} in order to compute absolute magnitudes of the dwarf candidates. The distance modulus uncertainty of 0.43 mag \citep{sorce2014dist} from the Tully-Fischer distance estimate corresponds to an 18\% uncertainty in the quoted distance. This does not affect our background selection, instead only shifts the range of the parameters that are derived in this paper based on the assumed distance, such as absolute magnitude, physical size and luminosity/mass.
    
    \section{Description of the Data}\label{sec:2}

\begin{figure*}
 \centering
 \includegraphics[width=1.2\textwidth,center]{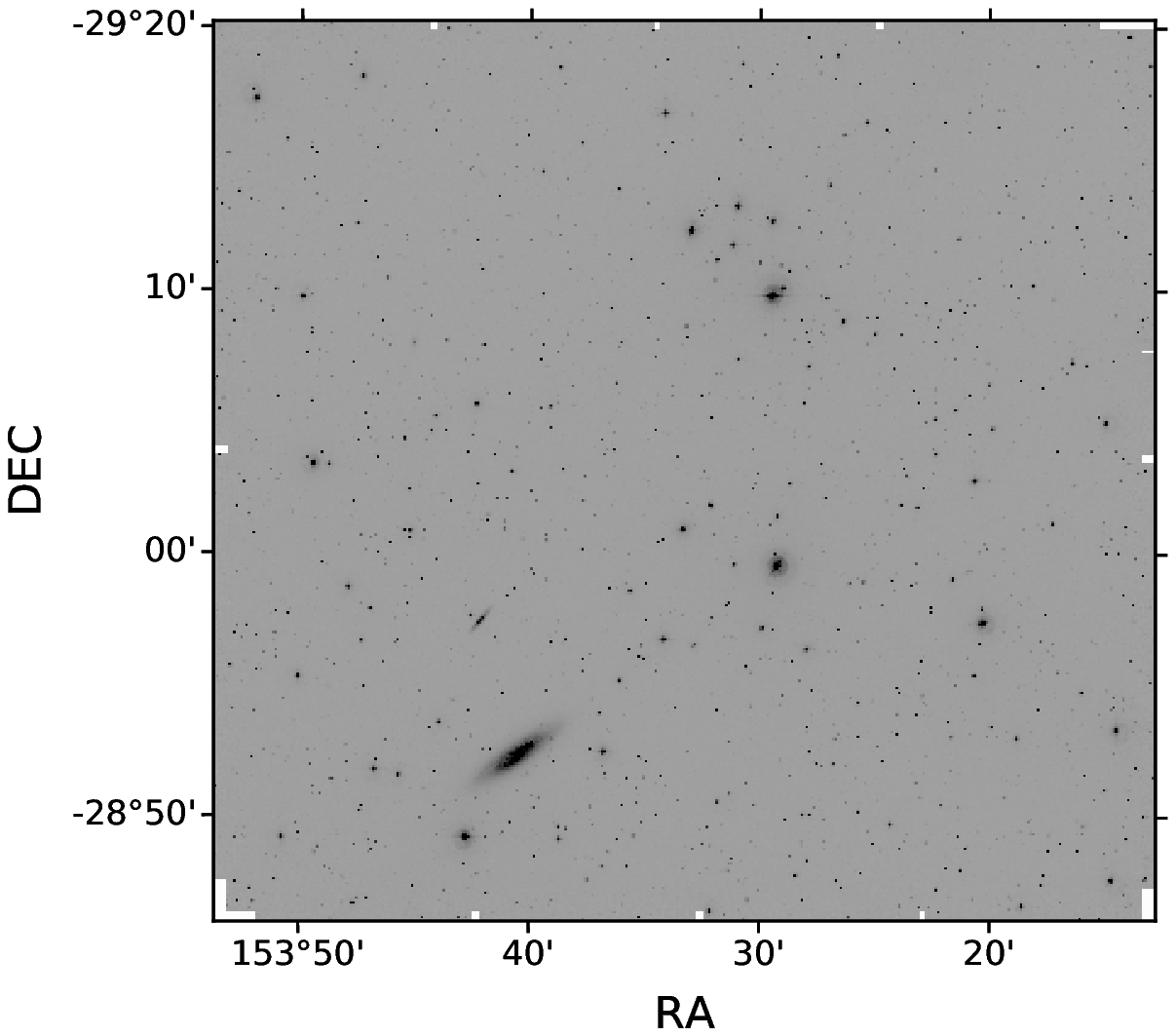}
\caption{The first field, F1R, with a field of view of 36$' \times $34$'$ showing NGC-3175 at the left corner.}
\label{fig:f1_wcs}
\end{figure*}

 	The imaging data used in this paper were obtained from the European Southern Observatory (ESO) Wide-Field Imager (WFI) instrument, mounted at the Cassegrain focus on the 2.2m Max Planck-ESO (MPG) telescope at the La Silla observatory, Chile \citep{baade1999wide}. The WFI has a large field of view of 34$' \times$33$'$ and a detector arrangement consisting of 4$\times$2 mosaic of 2k$\times$4k CCD's with a pixel scale of 0.238$''$/pixel. Imaging data was acquired in the broadband  R ($\lambda_{central}$ /FWHM:651.725nm/162.184nm) and B (451.0nm/133.5nm) bands in visitor mode between 30/01/2011 and 06/02/2011. The images were taken in 7 different fields to cover the central $\sim$500kpc of the NGC-3175 group. For each of the 7 pointings, multiple dithered exposures (typically 15) with an exposure time of 400s or 600s each (see Table~\ref{tab:im_prop}) were taken. This allows our imaging data to have total integration times of typically 9000s ($\sim$2.5hours) and to be 95\% complete down to M\textsubscript{B} $\sim$-8. As such, the properties of the combined image for each field varies and are summarised in Table~\ref{tab:im_prop}. We show the R band field with the NGC-3175 in Figure~\ref{fig:f1_wcs}.
 
    Prior to the analysis, the raw images were processed which included bias subtraction, flat field correction and registration to produce the 7 fields described in Table~\ref{tab:im_prop}. Overscan and trim methods were utilised to remove readout bias, whilst dark frames taken during the observation period were used to remove thermal electron counts from each CCD in the WFI array. The registration was performed to a common WCS using the NASA/IPAC UNSO B1 catalogue \citep{Monet2003UNSO}.

\begin{table}
	\centering
	\caption{Properties of the imaging data of the 7 fields covering the central regions of NGC-3175, acquired from the ESO WFI instrument. Exposure column indicates the number of dithered exposures combined to make the final image in R and B bands for a given field. The final column indicates the FWHM of the seeing estimated for each band used for star-galaxy separation. All B band images have an exposure time of 600s and the * indicates fields for which the exposure time in the R band is 400s.}
	\label{tab:im_prop}
	\begin{tabular}{lcr}
    \hline
Field & Exposures (R,B) & Seeing FWHM (R,B)\\ 
{} & {} & (arcsec) \\ \hline
F1 & 15,17 & 1.14,0.95 \\ 
F2 & 15,10 & 0.97,0.98 \\
F3 & 15,15 & 1.12,1.24 \\
F4 & 15*,15 & 1.70,1.43 \\
F5 & 14,15 & 0.95,1.06 \\
F6 & 15,15 & 1.02,0.94\\
F7 & 3,15 & 1.26,0.98 \\ \hline
	\end{tabular}
\end{table}

\section{Galaxy Detection and Star Removal}\label{sec:3}
	
    Detection of objects in our images is done by Source Extractor (SExtractor; \citealt{bertin1996sextractor}). SExtractor is an algorithm which automatically detects, deblends and performs astrometry and photometry of objects, optimised for speed and large images. We use SExtractor in ``dual band mode'', which allows us to specify a detection and a measurement image. The former image is used for detection and deblending purposes, whilst the latter is used for photometry. The R band is used as the detection image as it is the deepest image in every field, and photometry is done in both of the imaging bands. Such a method ensures that the same sized aperture is used to measure the magnitude in both bands, enabling an accurate measure of the colour.

	To run SExtractor in dual mode, we require the R and B band images to be registered and aligned. We work in pixel coordinates, and compute the geometric transformation required to map the R band coordinates to that of B band by fitting a power series to the 5\textsuperscript{th} order in x and y which accounts for x and y shifts, x and y scale factor and a rotation. We obtained accurate positions of the same set of stars that are non-saturated in both bands. Stars spanning the entire image are used to increase the accuracy of the alignment. To check alignment, we compute the centroids of the reference stars in both the R and B aligned images, finding that our best fields are aligned to within 0.01$''$ and the worst to within 0.05$''$, corresponding to the image with the worst seeing, F4. 

\subsection{Object Detection}\label{ch3:obj_det}
	Within an astronomical image the flux measured of a given pixel is the sum of the background (sky) intensity and the object's intensity. As we aim to detect the faint dwarf galaxy population in our imaging data, we require good background estimation. In the following sections, we describe in detail the choice of the main SExtractor parameters that affect the background determination and object detection. The first step of SExtractor is background measurement based on a method of $\kappa . \sigma$ clipping, where values that deviate more than $\pm 3\sigma$ from the mean are removed and the mean (for non-crowded fields) or mode (for crowded fields) of the remaining pixels is used to estimate the background \citep{bertin1996sextractor}.
    
	We choose the size of the background grid across which the background value is interpolated to be larger than the largest objects in the image. This ensures that small scale variations in the background are accounted for while not overestimating the background due to flux from objects.

	After the background subtraction process, we use a Gaussian filter to smooth the image as this filter is found to be best for the detection of faint objects.

\begin{table}
\begin{center}
\caption{Summary of the SExtractor parameters tuned for the detection of faint objects within our fields.}
\label{tab:sex_par}
\begin{tabular}{lr} \hline 
{Parameter Name} &  {Value} \\ \hline
BACK\_FILTERSIZE & {3 $\times$ 3} \\
BACK\_SIZE & 256 \\
FILTER\_NAME & gauss\_3.0\_7x7.conv \\
DETECT\_THRESH & 6$\sigma$ \\
DEBLEND\_NTHRESH & 16 \\ 
DEBLEND\_MINCONT & 0.05 \\ \hline
\end{tabular} 
\end{center}
\end{table}


\begin{figure}
\centering
\includegraphics[width=\columnwidth]{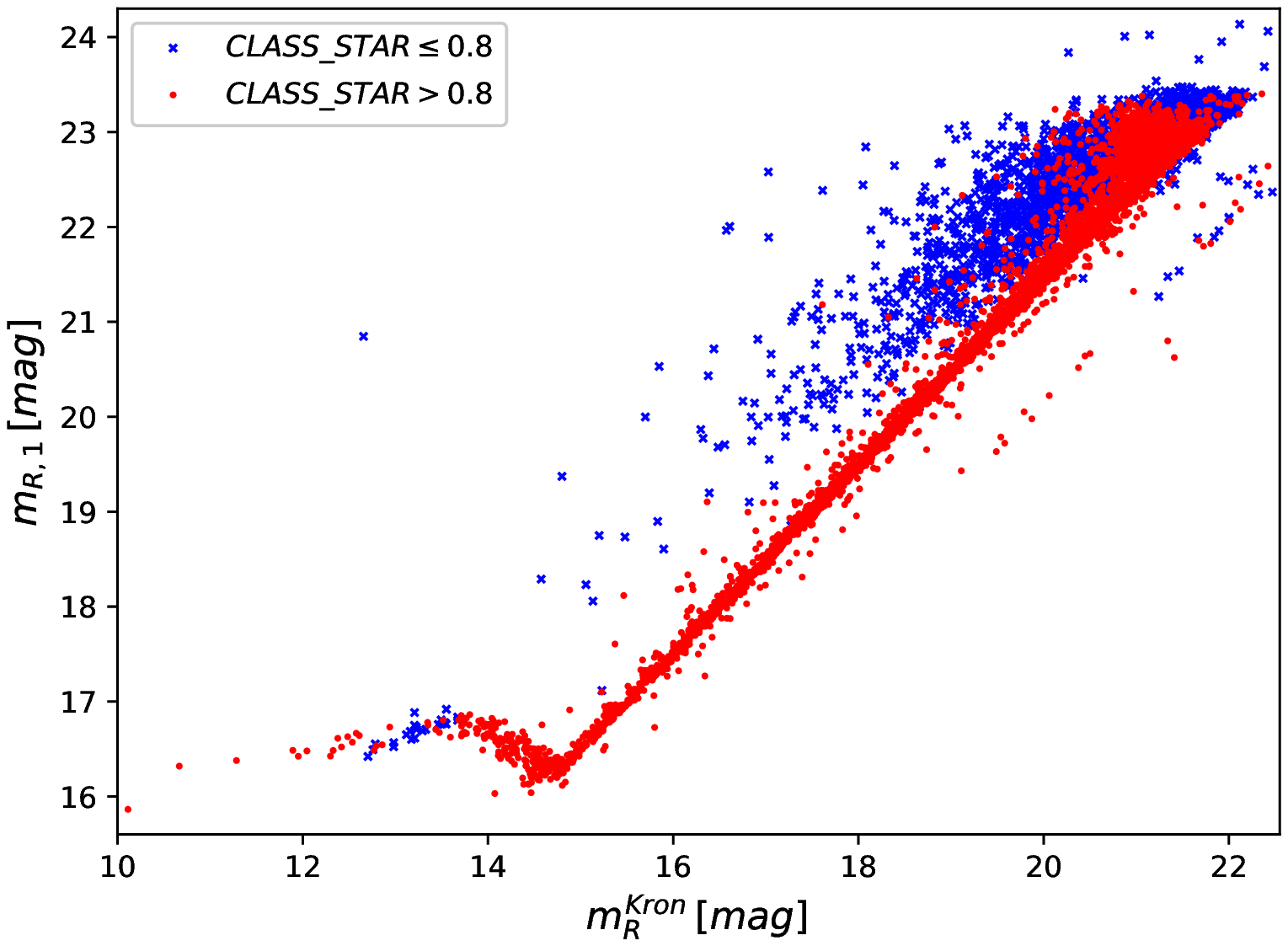}
\caption[1$''$ aperture magnitude against Kron Magnitude]{Central 1$''$ aperture magnitude against the SExtractor Kron magnitude in the R band, $m_{R}^{Kron}$ for F7. Objects classified as stars (\texttt{CLASS\_STAR} $>$ 0.8) are more centrally concentrated than galaxies and occupy a linear ``stellar locus''. The spread at fainter magnitudes is likely due to the limiting efficiency of the stellaricity index. Saturated stars occupy the bright tail end with large sizes.}
\label{fig:sat_star}
\end{figure}

The next SExtractor step is the detection of objects using a simple threshold process. A detection threshold of 6 $\sigma$ above the background was found to be ideal at detecting all of our faint objects while maintaining the number of spurious detections to $<$ 2\%, which were later manually removed.

    Accurate deblending of objects that appear nearby in sky projection is crucial for identifying sources to which the surface brightness profile needs to be fit (Section~\ref{sec:4}). The default deblending parameters were found to work best with our data, as listed in Table \ref{tab:sex_par}.
    A summary of other main SExtractor configuration parameters used in this study are also outlined in Table \ref{tab:sex_par}. 

    After the background estimation and object detection process, photometry on both the imaging bands is carried out using SExtractor by measuring the flux within an aperture. To determine a galaxy's total flux, we define the extent of a galaxy using the Kron radius, $R_{1}$ which defines the first moment of a galaxy as:
\begin{equation}
R_{1}(R) = \frac{2\pi \int_{0}^{R} I(x) x^{2} dx}{2\pi \int_{0}^{R} I(x) x dx}
\label{eq:kron_rad} \end{equation}
	where x is the radius. We use an aperture of size 2.5$R_{1}$ (defined using the R band) to measure the flux of an object. Such a size has been shown to theoretically contain more than 90\% of an object flux for S\'ersic indices ranging from 0.2 to 10 \citep{graham2005concise}. Magnitudes measured within this aperture are known as Kron Magnitudes.


\subsection{Star-Galaxy Separation}\label{sec:stargal}
	To identify stars, we use SExtractor's neural network based star galaxy separator, stellaricity index (\texttt{CLASS\_STAR}). The stellaricity index can range from 0 for extended objects, to 1 for point-like sources and requires an estimate of the seeing and the pixel scale of the image to be computed. We calculate the seeing of our images by measuring the PSF from isolated non-saturated stars. The PSF is measured as the FWHM of the distribution of light within these stars. We take multiple measurements across each image and average them to get an estimate of the seeing for each particular image.  The variation across each image in the PSF size is however negligible. We also use the stellaricity index from SExtractor to identify stars. This is a procedure in SExtractor which determines whether objects are extended or consistent with a point source. 

	Using a very high value of stellaricity index likely produces a high completeness of galaxies but also includes a large contamination of stars which would have to be manually removed, and vice versa. Furthermore, the robustness of the stellaricity index decreases dramatically at faint magnitudes \citep{chen2002campanas,bertin1996sextractor} and, when the sizes of objects are less than or comparable to the FWHM of the PSF. For the removal of stars, we perform various stellaricity index cuts and visually inspect any misclassified stars or galaxies. Via this process, we find that using a stellaricity index $>$ 0.8 provides a good compromise between achieving high completeness (see Section~\ref{fit_quality}) and purity of objects classified as galaxies. 

	Figure~\ref{fig:sat_star} shows a plot of the central 1$''$ aperture magnitude against the Kron magnitude, where our chosen stellaricity index cut identifies a linear stellar locus with a bright end tail corresponding to saturated stars. All objects that lie within 4 $\times$ isophotal area of such saturated stars are then removed from further analysis due to contamination from saturation spikes, and reflection of the telescope field corrector.
    
	We also remove objects with SExtractor $R_{e} < $ 1.4$\arcsec$ ($\sim$ 95 parsec) as these objects are likely background or compact stellar systems (such as globular clusters and UCDs; \cite{conselice2002galaxy,norris2014aimss, gregg2009ucd,liu2015ucd}). This will miss out any compact dwarf galaxies, however such objects are very rare and mostly found in rich clusters \citep{penny2012perseus}.
    
	Finally, the reduced catalogue obtained after the aforementioned star removal and size cut criteria for all our fields is visually inspected. In this process, we manually add back any missed galaxies without using any selection criteria based on size or structural/morphological features so as to avoid excluding possible interesting objects from further analysis. It is also found that usually $<$1.5\% of the objects in the reduced catalogue are stars misclassified as galaxies, however, this number rises to $\sim$ 10\% for one of the fields.

    The end result of repeating the first pass through SExtractor on all of our fields is a catalogue of 2574 potential dwarf candidates which are to be model fitted to calculate accurate shape parameters. As we show later, this number is $\sim$ 5 times larger than the final catalogue of dwarf candidates, from which we can infer that the first pass through SExtractor is unlikely to have missed a significant fraction of dwarf candidates.
    
\section{Galaxy Model Fitting}\label{sec:4}
	Photometry from SExtractor such as Kron magnitudes, calculated within the Kron aperture can miss some fraction of the flux of our objects, as mentioned in Section~\ref{sec:3}, affecting the estimation of shape and ellipse parameters of a galaxy. One must perform two-dimensional light profile fitting to determine accurate structural properties of galaxies. Many algorithms such as GIM2D \citep{simard2002deep} and BUDDA \citep{de2004budda} exist for this task, the most versatile and commonly used of which is GALFIT \citep{peng2002detailed}. GALFIT offers the ability of simultaneous fitting of multiple objects within the image and computes the best fitting model by a method of least squares minimisation, where the goodness of the fit is determined by computing the $\chi^{2}_\nu$ as
\begin{equation}
\chi^{2}_{\nu} = \frac{1}{N_{dof}} \sum\limits_{x=1}^{nx} \sum\limits_{y=1}^{ny} \frac{(f_{data}(x,y) - f_{model}(x,y))}{\sigma(x,y)^{2}}
\label{eq:chi}
\end{equation}
	where $N_{dof}$ is the number of degrees of freedom, $f_{data}(x,y)$ is the original image and $f_{model}(x,y)$ is the model image generated by GALFIT and, $x$ and $y$ correspond to each pixel in the image for which the calculation is summed over all $nx$ and $ny$ pixels.
    
\subsection{Automated Fitting Using GALAPAGOS-2}\label{section:automation}
    
    The task of manually fitting the 2754 objects identified in all of our fields, while adequately accounting for flux contribution due to nearby sources in a consistent method, is a very difficult task. We thus make use of GALAPAGOS-2 \citep{haussler2013megamorph}, an IDL routine which automates the process of running SExtractor, cutting postage stamps for the potential dwarf candidates, performs sky estimation and performs multi-band fitting using GALFIT-M (multi-band version of GALFIT). For a full description of the functional blocks of GALAPAGOS-2, we refer the reader to the release paper \citep{haussler2013megamorph} and the MegaMorph project.\footnote{http://www.nottingham.ac.uk/astronomy/megamorph/}

    A target list of objects to be fitted can be supplied to GALAPAGOS-2. Objects are fit in descending order of their magnitudes (i.e. brightest first) to ensure that flux contribution of the brightest objects is properly accounted \citep{haussler2013megamorph}. Therefore, for every potential dwarf candidate, we add any objects (detected by SExtractor) that are brighter and within 150$\arcsec$ of the candidate, to the target list to be fit.

	A major consideration for the fitting routine is the PSF used for convolution with the model which is crucial for accurate photometry. An ideal PSF would have a very high signal-to-noise ratio (S/N) with a flat and zero background \citep{peng2002detailed}. We build a PSF by using a sample of non-saturated stars, requiring that all have a total S/N $>$20. All PSFs generated by this method are by default centred at the peak flux and have a zero background. To test the PSF, we inspect the residuals after the PSF is fit to different stars. We use the empirical PSF which have a variety of FWHM values in the different pointings due to the seeing. The sample of stars used to build the PSF are varied to reduce the $\chi^{2}_{\nu}$ goodness of fit.

\begin{table}
\begin{center}
\begin{tabular}{lccr} \hline 
{Parameter Name} &  {a}  & {c} & {Error} \\ \hline
$m_R$ & 0.0043 & 0.0148 & 0.1 (mag)\\
$R_e$ & 0.064 & -1.46 & 0.4$''$\\
$n$ & -0.0196 & 0.2987 & 0.13\\
$\frac{b}{a}$ & -0.0018 & 0.0312 & 0.008\\ 
$m_B$ & 0.0043 & 0.0203 & 0.1 (mag) \\ \hline
\end{tabular} 
\caption{Coefficients for the line of best fit (Equation~\ref{eq:sim_er}) of deviations of the important fitted parameters.}
\label{tab:errors}
\end{center}
\end{table}

\begin{figure*}
\centering
\includegraphics[width=\textwidth]{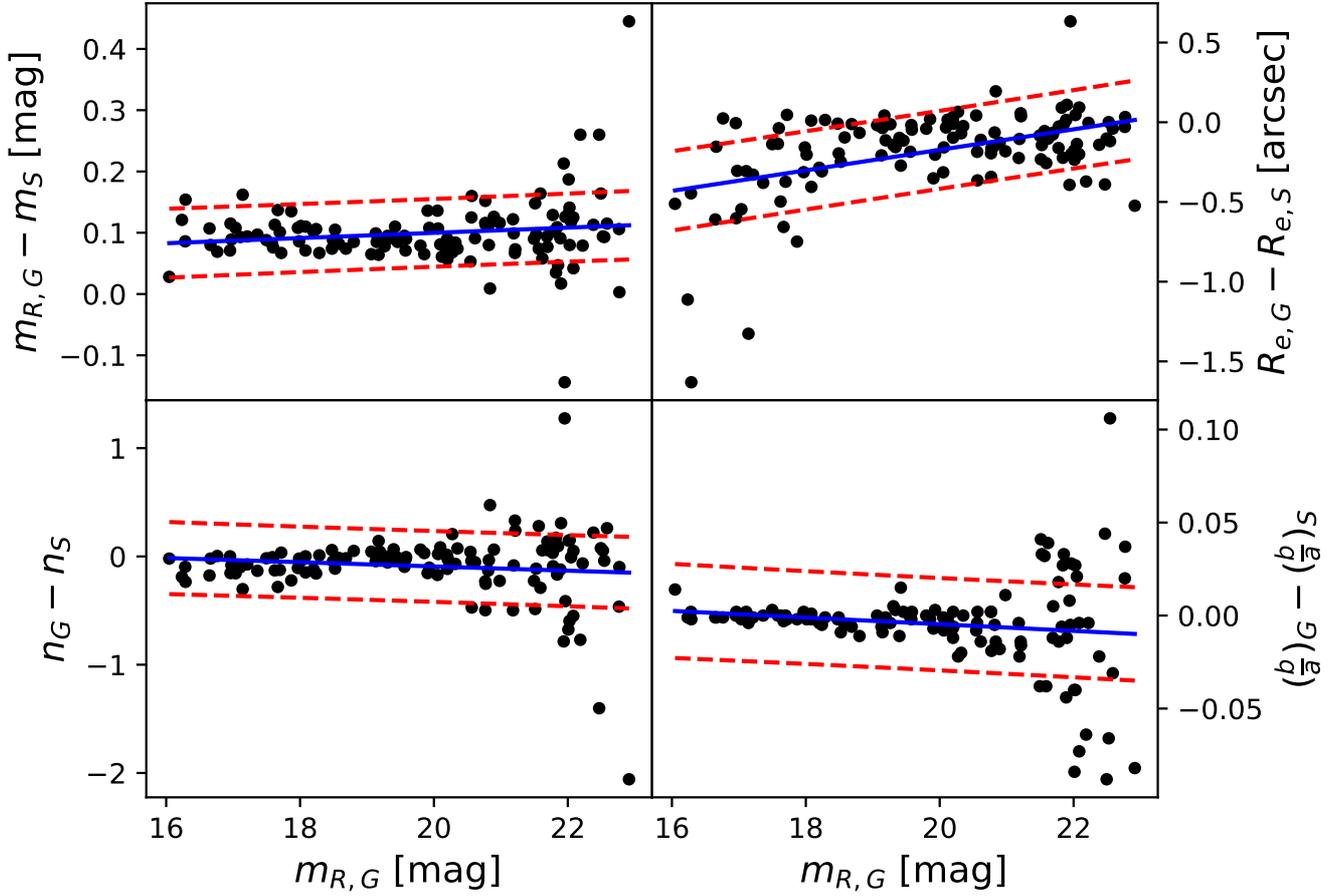}
\caption[Simulated Image Errors]{Analysing the image simulation output using the deviation between GALFIT and simulated values for each major parameter. The deviation for each free parameter is calculated and plotted on the y-axis according to Equation~\ref{eq:sim_er} as a function of the R band fitted magnitude. The subscripts G and S stand for GALFIT output and (initial) simulated values, respectively. The line of best fit, with 1$\sigma$ confidence intervals are also included on each plot. The overall low scatter reinforces the efficacy of the galaxy fitting methodology (see Table~\ref{tab:errors}).}
\label{fig:sim_err}
\end{figure*}

	One of the most common parametric functions used to study light distribution and galaxy morphologies is the S\'ersic profile, with the form
\begin{equation}
I(R) = I_{e} exp \left\lbrace -b_{n} \left[ \left( \frac{R}{R_{e}} \right)^{1/n} - 1 \right] \right\rbrace
\label{eq:sersic}
\end{equation}
    where $I_{e}$ is the intensity at the effective radius $R_{e}$, $n$ is the S\'ersic index and $b_{n}$ is a coefficient dependent on the S\'ersic index \citep{graham2005concise}. The S\'ersic index describes the shape of the profile where the $n$=1 case corresponds to an exponential light profile and the $n$=4 case corresponds to the de Vaucouleurs profile.

	We fit a single component S\'ersic model (Equation~\ref{eq:sersic}) to each of the target list objects. The free parameters for the fit are the position [$x$, $y$], total magnitude $m$, effective radius $R_e$, the S\'ersic index $n$, the axis ratio $q$ ($q = b/a$; semi-minor to semi-major half-axis ratio) and the position angle $\theta$. For the multi-band fitting feature of GALFIT-M, we constrain the free parameters such that only the total magnitude of an object can vary between the R and B bands, providing a more robust fit compared to single band fitting. We also perform a GALFIT test run on one of the fields while allowing the structural parameters to vary between the two bands during the fitting process. As structural properties of galaxies can vary between the two bands, this test run was performed to check if this would have significant effects on the fitted parameters when compared with a run where structural parameters were fixed between the two bands. We find negligible difference in the fitted parameters between the two runs, with significant deviations occurring generally for sources with a high $\chi^{2}_{\nu}$.

\subsection{Image Simulations}\label{im_sim}
	GALAPAGOS-2 also outputs uncertainties for each of the free parameters fit to the model. However, the uncertainties calculated by GALFIT are based on Poisson noise statistics and hence are incredibly small and (some debate) even meaningless \citep{peng2002detailed}. To determine the quality of the fitting and to determine an estimate of the uncertainties of the fitted parameters, we create simulated data on which we run all of our previous processing steps from SExtractor to GALAPAGOS-2.

	The properties of the simulated sample of galaxies were derived using the fits of the dwarf candidates in our fields. The simulated galaxies have magnitudes in R ranging from 16 to 22.5 mag, $\sim$ 0.5mag fainter than the magnitude limit of our data to allow us to test the completeness of our methods. The galaxies are distributed within this range using a Schechter luminosity function with a power law exponent, $\alpha = -1.24$ to obtain a steep luminosity function at the faint end, populated by dwarf galaxies. We only simulate galaxies with S\'ersic indices of either $n=1$ or $n=4$. The effective radii of the galaxies are scaled according to the size of the faintest galaxy which, we estimate to be $R_{e} \sim$ 0.7$\arcsec$ by extrapolating the effective radius - magnitude relationship of our fitted objects. The remaining free parameters, axis ratio and position angle are uniformly distributed between 0 to 1 and 0\degr to 359\degr, respectively.
    
    We first make a blank image using the sky, gain and noise properties of one of our fields (F2) to simulate a realistic mimic of the background of our imaging fields. We use a uniform spatial distribution to insert the 120 galaxies with the above properties on the simulated image. We employ a similar method for making a ``B band image'' for the exact same galaxies by using the median (B - R) colour of real galaxies in our sample to set the magnitude in B band. The PSF is assumed constant across each image, using the PSF from F2 as a template.
    
	To assess the completeness of our method of generating a target list for GALAPAGOS and the quality of the fit for GALFIT, we perform all of the aforementioned analysis in Section~\ref{sec:3} (galaxy detection) and Section~\ref{section:automation} (galaxy fitting) on the 120 simulated dwarfs. The first pass through SExtractor detects 118/120 galaxies. Implementing the automated dwarf candidate selection criteria described in Section~\ref{sec:stargal}, we obtain a catalogue of 63 potential dwarf candidates, corresponding to 53\% of the detected (by SExtractor) galaxies. The remaining 47\% of the galaxies can be split into two groups. Firstly, there are 26 galaxies that are fainter than the magnitude limit of our real data. As our object detection and extraction method is fine tuned to the magnitude and size limit of our imaging data, it is not surprising that those 26 galaxies are not included as potential dwarf candidates by the criteria defined in Section~\ref{sec:stargal}. Secondly, there are 29 galaxies remaining that are not identified as potential dwarf candidates by our automatic dwarf galaxy selection method. We therefore use this to estimate the completeness of our automatic dwarf galaxy selection method at $\sim$ 75\%. However, as mentioned in Section~\ref{sec:stargal}, we visually inspect all of the real images after this automatic selection of galaxies and, we manually add any dwarf candidates that are missed by our selection process.

\begin{figure}
\begin{center}
\includegraphics[width=1.2\columnwidth,center]{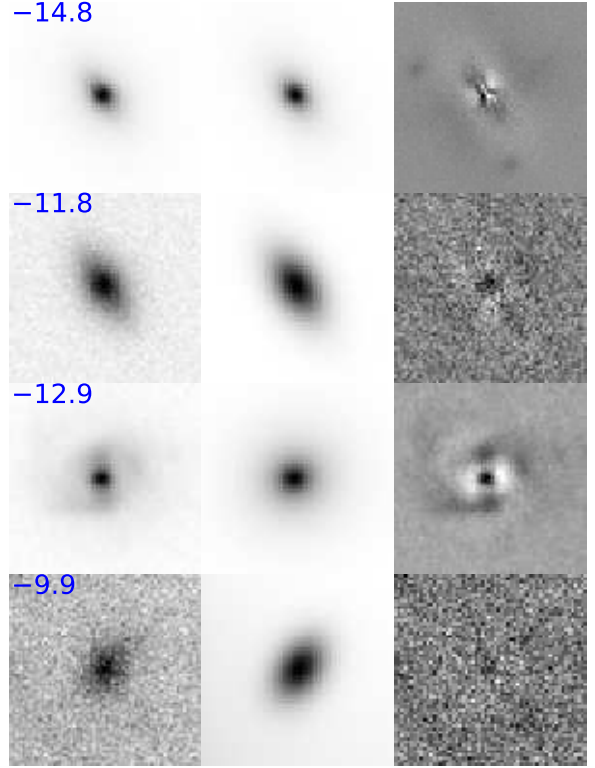}
\caption[Examples of model fitting, including a spiral]{A small set of the galaxy images that are fitted (\textit{left}), alongside their respective models (\textit{middle}) and residuals (\textit{right}). \textit{Third row} especially, shows that we are able to identify spiral galaxies with the resolution of our images. The absolute R-band magnitude is printed atop the left tile of each row.}
\label{fig:spiralexample}
\end{center}
\end{figure}

\subsubsection{Quality of the Fits}\label{fit_quality}
	All of the 118 simulated galaxies are subsequently fitted by GALAPAGOS. By comparing the properties of the simulated galaxies with that of the model fit, one can estimate the uncertainties for each of the free parameters in the model. We show in Figure~\ref{fig:sim_err}, the deviations of the fitted parameters from the true values, as a function of the fitted R band magnitude. The linear least squares regression fit between the deviation of a parameter and the magnitude, along with the corresponding 1$\sigma$ confidence intervals are also displayed. The linear fit to the deviation is calculated in the form
    \begin{equation}
    \left( Fit_{x} - Sim_{X} \right) = a\times m_{R,G} + c
    \label{eq:sim_er}
    \end{equation}
	where $Fit_{x} - Sim_{X}$ gives the deviation between the fitted parameter values and the simulated values, computed for each free parameter. $m_{R,G}$ is the fitted R band magnitude, and $a$ and $c$ are the gradient and intercept of the line of best fit, respectively. Generally, the deviations in the parameters get worse at fainter magnitudes, and beyond the magnitude limit of our real images, as shown by the scatter in Figure~\ref{fig:sim_err}. However, the opposite is true for the effective radius deviation, which gets worse for brighter, larger objects.
    
    We use the gradient and intercept parameters of the line of best fit to estimate the uncertainty in our fitted parameters. These uncertainties are much more sensible than those calculated by GALFIT although they are likely a lower bound for the true uncertainty. This is because although we use the readout noise, gain and sky parameters to make an image, inserting galaxies onto each of the imaging fields would provide a more realistic estimate of the uncertainty as the any seeing effects or distortions in the real image will be included in the model fit and therefore the uncertainty estimate. In addition, this would also increase the sample size of the simulated galaxies, providing a more robust statistical measure of the deviations from the simulated (true) values. We did not insert the simulated galaxies onto our real images as the crowding becomes too high for our SExtractor configuration to work accurately, which then also affects the GALFIT process.

\begin{figure*}
\centering
\includegraphics[width=\textwidth]{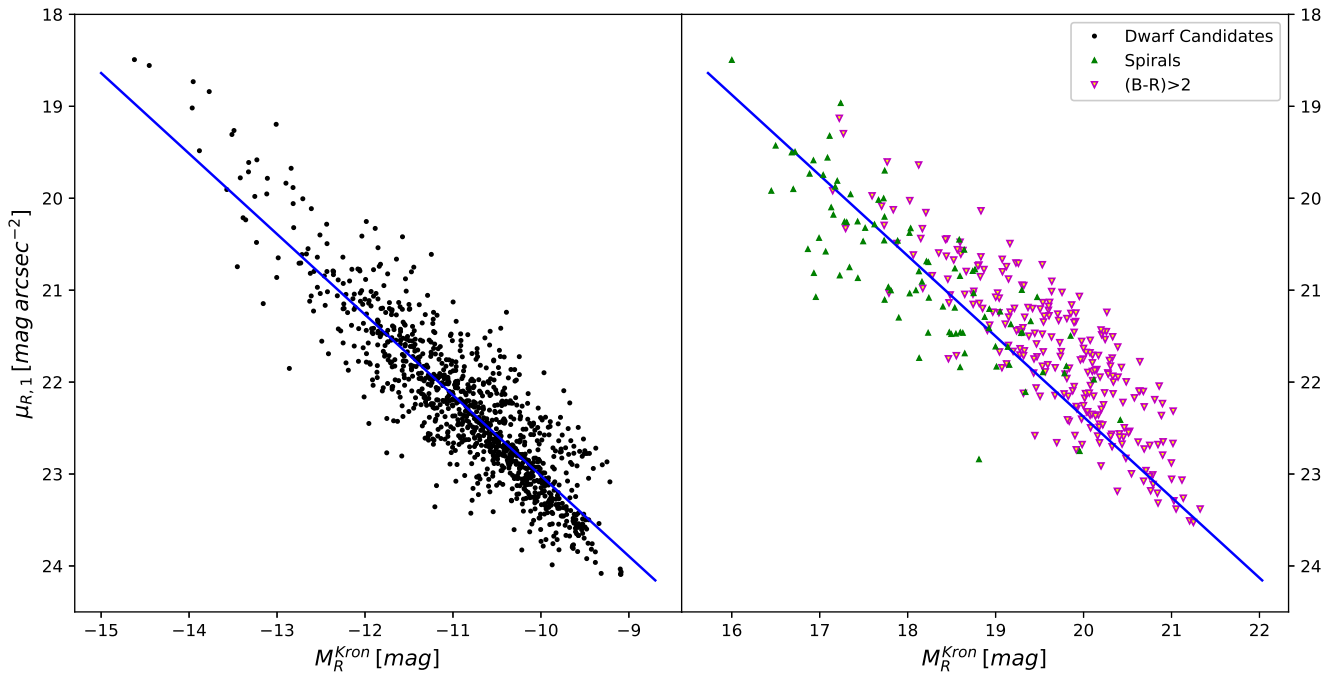}
\caption[Central surface brightness against Kron Magnitudes]{\textit{Left:} Plot of central surface brightness (within 1$''$ aperture) against inferred absolute magnitude $M_R$ (both from SExtractor) for the dwarf candidates following initial background removal using colour and visual morphological criteria. The linear least squares fit to the data is also plotted. Background galaxies appear further away from the best fit line and hence objects 1$\sigma$ away from this relationship are removed. \textit{Right}: The same relationship and line of best fit but as a function of Kron apparent magnitude $m_{R}^{Kron}$ for objects that are visually identified as spirals (green) and objects that are very red (pink). These objects show larger scatter about the best fit line, as would be expected for background galaxies.}
\label{fig:csb_bkg}
\end{figure*}

\subsection{Removal of Background Galaxies}\label{sec:bkg_gals}
	In the absence of spectroscopic data, we employ three criteria based on the morphological and photometric properties of the galaxies in our sample to remove contamination from background galaxies. Firstly, we apply a colour cut such that objects redder than (B-R) $=$ 2 are classified as likely background galaxies \citep{conselice2002galaxy} although, we note that such a cut will also remove any dwarf galaxy members with peculiar stellar populations but these are likely very rare. A colour criteria based on the same principle is found to identify background galaxies in the core of the Virgo cluster \citep{lieder2012deep}.

	Extensive studies of dwarf galaxies in the Local Group and clusters (e.g., \citealt{binggeli1985virgo,conselice2002galaxy,2012mcconn}) have found that dwarf galaxies, especially near the central regions, generally have a symmetric elliptical/circular shape and are smooth with no internal structure such as dSphs and dEs. In contrast, background galaxies, especially in the field, can usually be identified visually due to internal structure such as spiral arms, which are visible and resolved at the resolution and depth of our imaging data. This is evident in Figure~\ref{fig:spiralexample} (\textit{third row}) which shows the model fit process - input galaxy, GALFIT model and the residuals along each row. Therefore, we study the structural properties by visual inspection of the model and residuals of all the dwarf candidates to remove spirals as likely background galaxies based on the above defined criteria.
    
    Galaxies with asymmetric components such as bulge/disk are not fitted well using a single S\'ersic component. This is evident in the residuals and such objects are hence removed from the catalogue as the fitted parameters are likely not reliable.
    
	Then, we refit ``by hand'' objects with high $\chi^{2}_{\nu}$ or bad residuals, usually due to the presence of very bright or large stars in the fitting region that are generally not modelled well by the PSF. Such nearby stars were manually masked, and the object refitted before further classification and analysis. Objects which continued to have a bad fit were subsequently removed from further study, as the fitted parameters are likely not accurate. Approximately 22\% of the objects were thrown out due to bad fitting. Some of these objects, especially those with highly asymmetric structure are likely to be background systems given their complex morphology as most cluster dwarfs are well fit by a S\'ersic profile. 

	Finally, we study the apparent and absolute size of the galaxies in our sample to constrain the dwarf candidates further. We plot the tight relationship between he central 1$''$ surface brightness ($\mu_{R,1''}$) and absolute magnitude ($M_{R}^{Kron}$) and perform a linear least squares fit to this relationship (blue line; Figure~\ref{fig:csb_bkg} \textit{left}) and obtain
\begin{equation}
\mu_{R,1''} = 0.87 M_{R}^{Kron} + 31.77
\end{equation}
to be the best fitting parameters. Surface brightness (SB) is independent of distance for non-cosmological distances (which is true for our group). Therefore, a background galaxy can be small in size, but still contain a high surface brightness. Hence, the $\mu_{R,1''}$ - $M_{R}^{Kron}$ relationship can be used to asses group membership of dwarf candidates. This relationship has been used by others (e.g., \citealt{conselice2002galaxy,muller2017centaurus,muller2017m101,muller2018leo}) to identify background galaxies, and in the case of \cite{conselice2002galaxy}, identified dwarf galaxies that were later confirmed with spectroscopy by \cite{Penny2008faint}. We remove objects which deviate significantly by more than $1\sigma$ from the line of best fit. The tight correlation of the relationship and the lack of clusters of points far from the best fit line suggests that our sample is mostly drawn from one distinct population - the galaxy group. This increases our confidence in the effectiveness of the background removal criteria employed.

\begin{figure*}
\centering
\includegraphics[width=\textwidth]{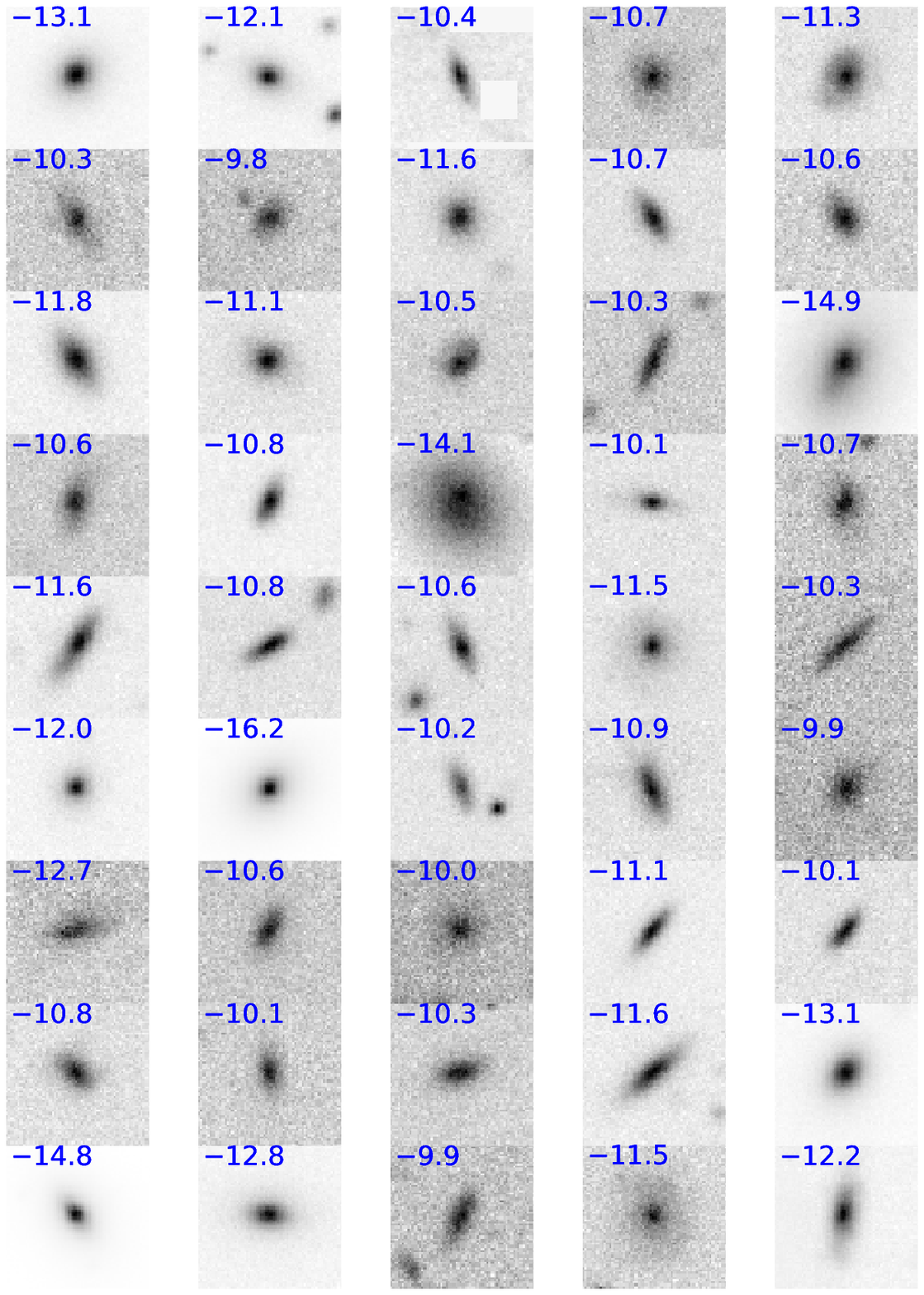}
\caption[Sample of Dwarf Candidates]{Representative sample of dwarf candidates taken from all of our fields with a length of $\sim 15''$ on each side. The absolute R band magnitudes are printed atop each image.}
\label{fig:out_gals}
\end{figure*}

	In Figure~\ref{fig:csb_bkg} (\textit{right}), the same line of best fit is shown but as a function of apparent magnitude for objects that were selected to be background by either the colour cut, or the morphological classification criteria. As evident, the scatter about the best fit line is larger than for the dwarf candidates as would be expected of background galaxies. Surprisingly, a large fraction of objects selected as background systems lie on/close to the best fit line. This may suggest the possible problems with colour based background selection, as discussed later in Section~\ref{sec:bkg_gals}. 
    
    Also important to note is that the distribution of dwarf candidates revealed no visible clustering, which further suggests that we do not select any background cluster members as dwarf candidates for the NGC-3175 group. In addition, we note that our selection criteria is as conservative as possible in the absence of spectroscopy.

    Our resulting catalogue after performing this morphological and colour based selection consists of 553 dwarf candidates of the group (hereafter, dwarf candidates). In comparison, the final dwarf candidates form $\sim$20\% of the 2574 initial potential dwarf candidates from first pass through SExtractor. We show a representative sample of the dwarf candidates in Figure~\ref{fig:out_gals}.
    
    However, there are some issues with using colour and morphology to define background galaxies. \cite{conselice2002galaxy} and others have found that there exists a signification population of background galaxies, identified by the presence of peculiar internal or spiral structure, with (B-R) colours $<$ 2. Hence, it is likely that there may still be some contamination from background galaxies in our sample, which is inevitable.
    
\section{Analysis}\label{sec:5}
\begin{figure*}
\centering
\includegraphics[width=0.9\textwidth]{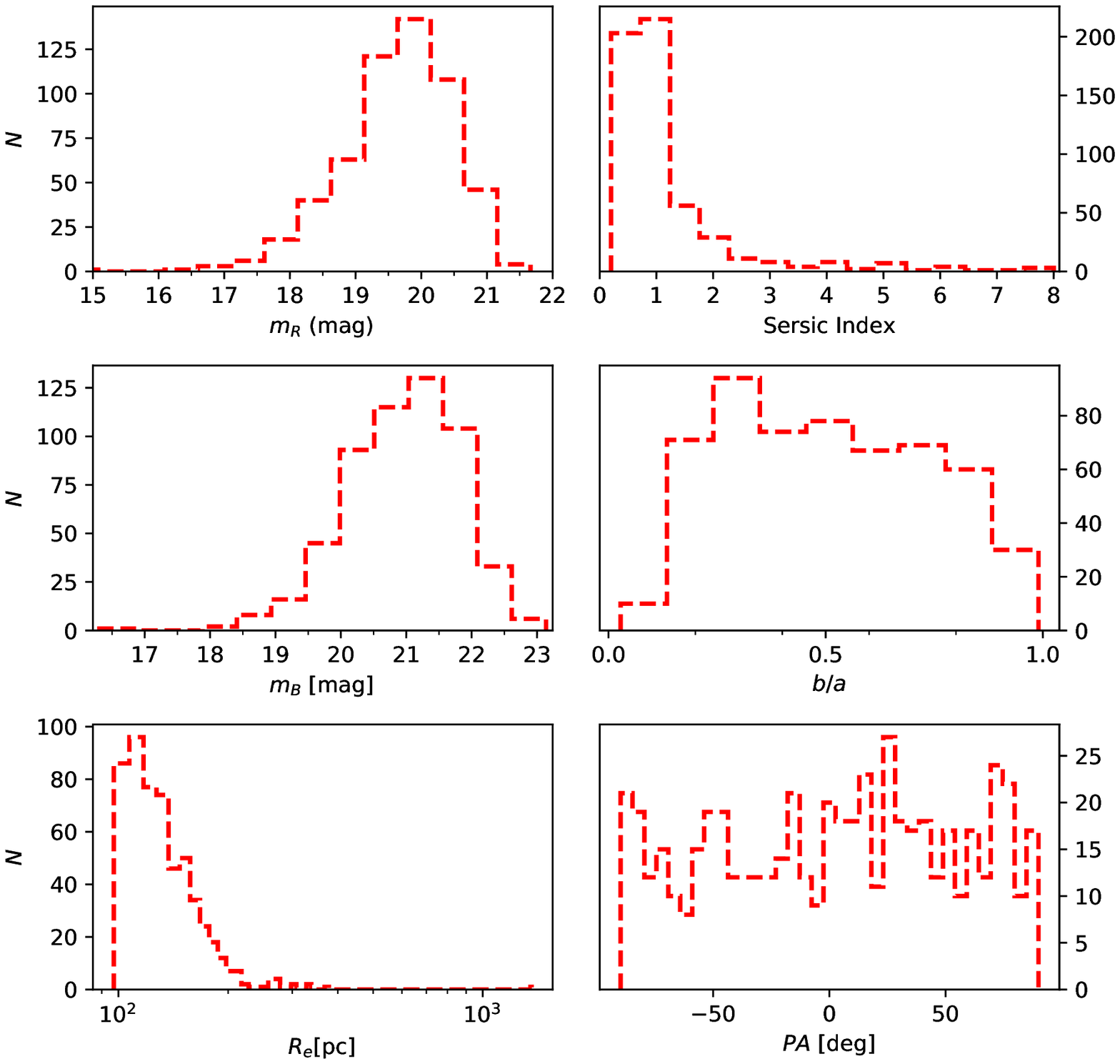}
\caption[Distributions of fitted parameters]{Distribution of fitted parameters for the dwarf candidates. These all align as expected for relatively small and faint dwarf galaxies, with S\'ersic indices distributed closely around 1 (exponential light profile).}
\label{fig:par_dist}
\end{figure*}

	We now investigate the properties of the identified dwarf candidates in the subsequent sections. By analysing which regions exactly these galaxies occupy in key plots such as magnitude-size, and surface brightness-magnitude relations, we can begin to further understand the satellite population of the NGC-3175 group. Prior to this, it is useful to inspect the distributions of each fitted parameter of the dwarf candidates from GALFIT. 
    
    Figure~\ref{fig:par_dist} shows the distributions of the parameters fit by GALFIT. Most significantly, most of our objects have S\'ersic indices close to $n$ = 1, i.e. follow exponential profiles, with a small fraction centred at a de Vaucouleurs profile (with $n$=4). As anticipated for a dwarf population, the objects are primarily skewed towards the smaller end for effective radii, and towards the fainter end in magnitude. The axis ratio $ \left( \frac{b}{a} \right)$ spans the entire interval, as do the position angles, with no dramatic spikes nor troughs in their distribution.

\begin{figure*}
\centering
\includegraphics[width=\textwidth]{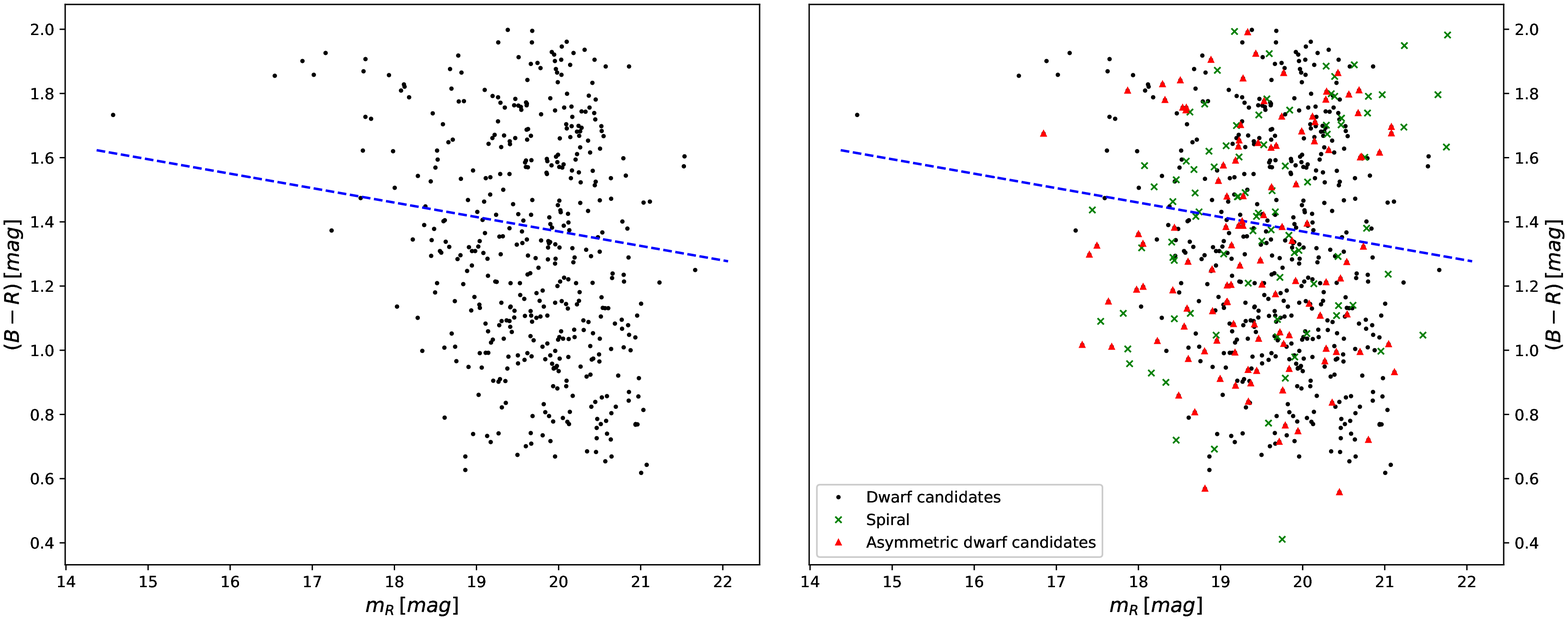}
\caption[Colour-Magnitude diagram for the Group Members]{Colour-Magnitude diagram for the distinct morphologies of galaxies in our sample along with the red sequence line observed in clusters \citep{adami2006a}. \textit{Left:} Only the dwarf candidates. \textit{Right:} Dwarf candidates, dwarf candidates visually identified to have asymmetric structure and objects with spiral/disky structure. The dwarf candidates identified are likely an analogue in colour-magnitude space to the Low Mass Cluster Galaxies (LMCGs) discovered in the Perseus Cluster by \cite{conselice2002galaxy}. Unlike in clusters, we do not observe a distinct red sequence in our group, possibly as we do not probe the bright region where the CMR is typically observed.}
\label{fig:colmag}
\end{figure*}

\subsection{Colour-Magnitude Relationship}\label{sec:colmag_rel}
    The colour-magnitude relationship (CMR) of dwarf galaxies has been observed and studied in detail, especially for nearby clusters such as Virgo (\citealt{binggeli1985virgo,lieder2012deep}), Coma \citep{adami2006a,yagi2016catalog} and Perseus \citep{conselice2002galaxy}. All find the presence of a red sequence, for the bright galaxies ($M_{B} < -16$), which identifies the large, passive galaxies. Figure~\ref{fig:colmag} shows the colour-magnitude diagram for our dwarf candidates (black dots) using the total (B-R) colour. For reference, we also show the dwarf candidates which appeared visually asymmetric (red triangles) and, the objects that were found to contain spiral/disky structure (green cross).
    
	Interestingly, we do not detect very many luminous galaxies in the range where the red sequence is typically fit. At intermediate to faint ($M_{R} >$ -12) magnitudes, this population of dwarf candidates appears to be an analogue in colour-magnitude space of the low mass cluster galaxies (LMCGs) population observed by \cite{conselice2002galaxy} in the Perseus cluster.

 \subsection{Absolute Magnitude - Effective Radius}\label{sec:M_vs_re}
 
\begin{figure*}
\centering
\includegraphics[width=\textwidth]{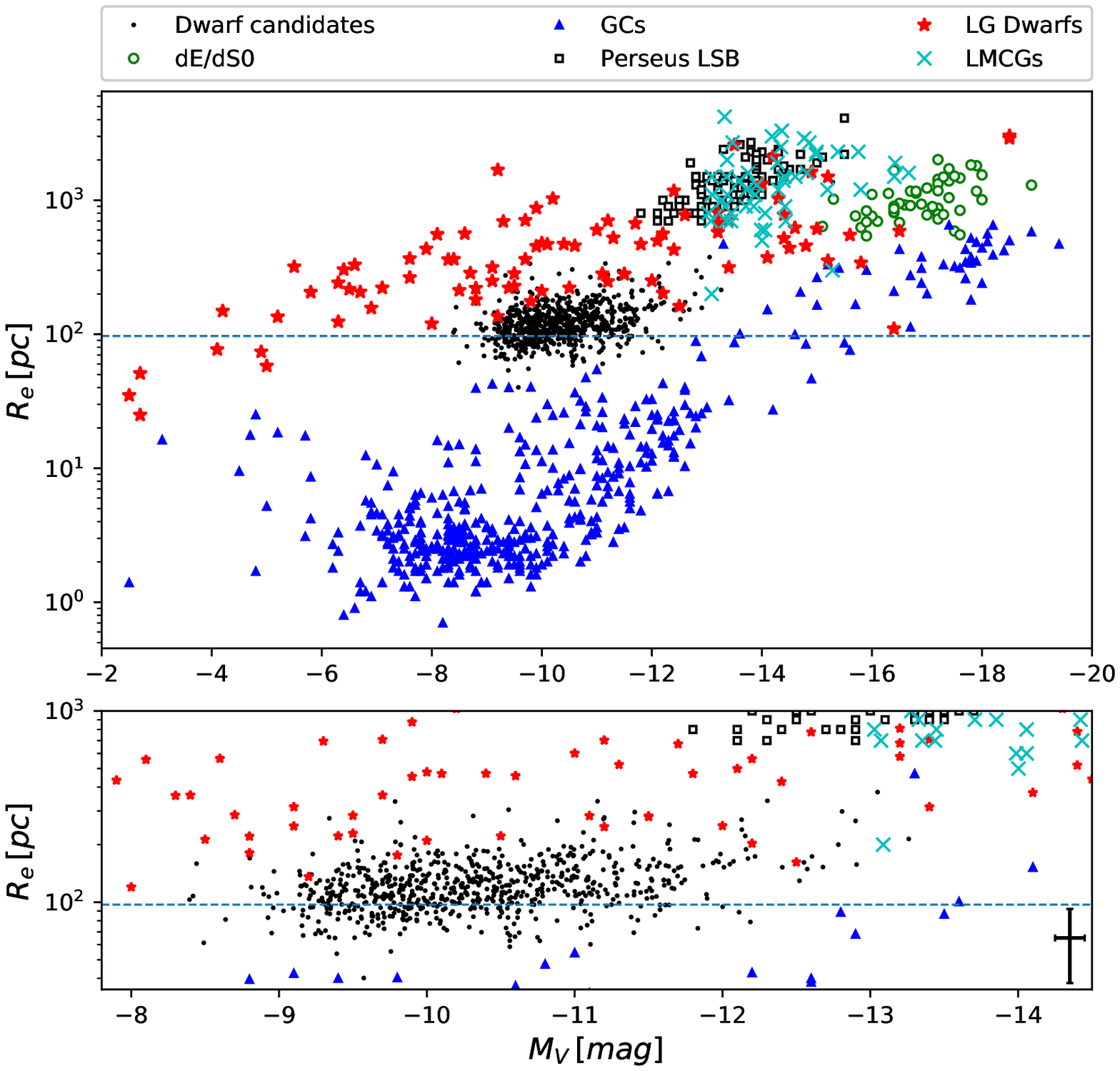}
\caption[Magnitude - Effective Radius relationship]{$M_v$ vs $R_{e}$ plot for our dwarf candidates (black dots) and literature data, for comparison. \textit{Bottom:} Scaled to focus on our candidate objects. This more clearly shows the slight overlap with a few GCs, and where the candidates fall close to the dSph/dE locus. Dwarf candidates of the NGC-3175 group are generally smaller than the dwarf galaxies found in both the MW and M31. The horizontal line indicates the initial size limit of our initial selection for the dwarf galaxies. Some galaxies fall below this line after a proper measurement of their sizes with GALFIT (see text). Also shown in the \textit{bottom} panel is the typical error bar for our dwarf galaxies (the location of the error bar does not represent a real data point). Literature data included on this plot: \footnotesize LG dwarfs (red stars): \cite{Walker2009dSphdE,2012mcconn,tollerud2012splash,tollerud2013m31satellites}, and references therein. dE/dS0 (green circles): \cite{geha2002virgo,geha2003rotating,Chili2009virgo,toloba2012virgo,forbes2011}, GC (blue triangles): \cite{brodie2011relations}, Perseus LSB (hollow squares): \cite{wittmann2017perseus}, LMCGs (cyan crosses): \cite{conselice2003perseus}.}
\label{fig:re_mv}
\end{figure*}

	In Figure~\ref{fig:re_mv}, we plot the effective radius of the dwarf candidates versus their absolute V band magnitude, and compare with other dwarf/stellar systems. We do not have V band imaging for our data so, we perform a linear interpolation between the R and B-band magnitudes to estimate the V band magnitude for our objects. The literature data of the dwarf and stellar systems are primarily taken from the catalogue compiled by \cite{norris2014aimss} and the references therein. We show the properties of the MW and M31 dSphs \citep{2012mcconn,Walker2009dSphdE,tollerud2012splash,tollerud2013m31satellites}, the dE/dS0 sample \citep{geha2002virgo,geha2003rotating,Chili2009virgo,toloba2012virgo,forbes2011}, the Low-Mass Cluster Galaxies (LMCGs; \citealt{conselice2003perseus}) and, the GCs \citep{brodie2011relations,strader2011aM31} in the size-magnitude plane. Even at our magnitude limit, it is possible to distinguish the GCs (and other compact stellar systems) from the dwarf candidates due to their distinct locations in the size-magnitude plane. The horizontal line shows the initial size limit (based on SExtractor size) employed on our data (see Section~\ref{sec:stargal}) to remove small objects such as globular clusters. However, after fitting with GALFIT, the sizes of some of the objects changed from their SExtractor measures. As a result some of these objects are now below our initial size limit after fitting. We also find that at the lower end of the size scale SExtractor over-estimates the sizes of galaxies from what we find with GALFIT.
	
    We find that our dwarf candidates (filled black circles) appear to follow the scaling relation similar to that of the larger and brighter dEs (hollow green circles) and are close to the dSph area of the size-magnitude relationship. This suggests that our candidates and previous dEs have a similar origin. We also note that our dwarf candidates are significantly smaller in size to the dSphs found in the MW and M31. Although, the fitted output of simulated galaxies (Section~\ref{sec:4}) revealed no systematic underestimation of sizes hence, this relationship is likely not due to our analysis methods. Furthermore, the dispersion in the relationship is small which suggests that our sample is a selection of the faint population and hence we may be missing some of the very large dwarf galaxies.   However, we are measuring the sizes of our objects very accurately based on our simulations, and therefore we are likely measuring sizes with a much smaller scatter than what can be done with Local Group galaxies which suffer from significant contamination due to foreground/background objects.
    
\begin{figure*}
\centering
\includegraphics[width=\textwidth]{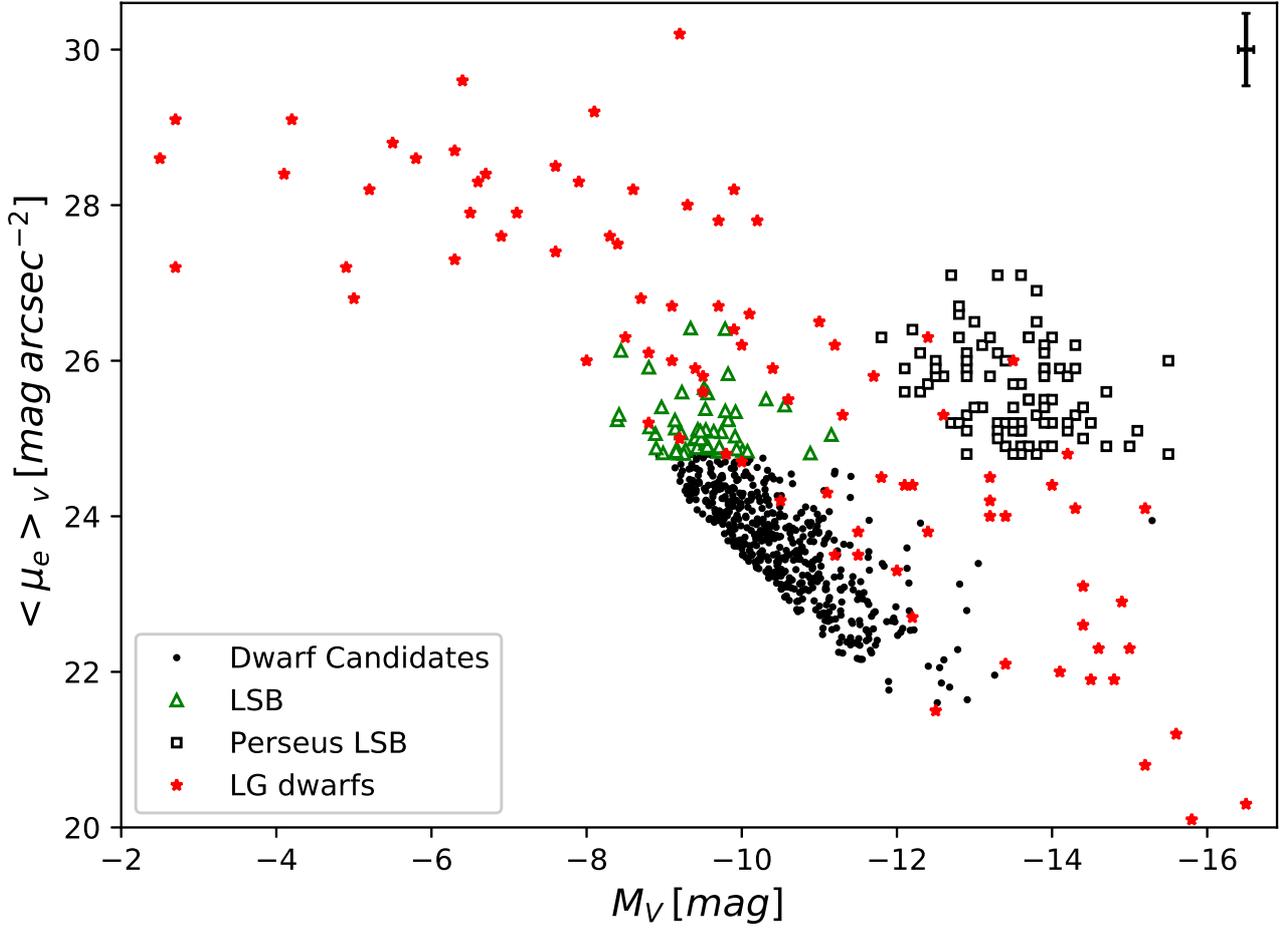}
\caption[Surface Brightness - Magnitude Relationship]{$\langle\mu_{e}\rangle_{v}$ versus M\textsubscript{V} for our dwarf candidates with Perseus LSB galaxies\citep{wittmann2017perseus} and LG dwarfs (\citealt{2012mcconn}, and references therein) plotted alongside. Additionally, we plot the 53 objects in our sample that meet the LSB criteria ($\langle \mu_{e} \rangle_{v} \geq $ 24.8 mag arcsec\textsuperscript{-2}) as green triangles. Also shown is the typical error bar for our dwarf galaxies (the location of the error bar does not represent a real data point).}
\label{fig:mueff_mv}
\end{figure*}

\begin{figure}
\centering
\includegraphics[width=\columnwidth]{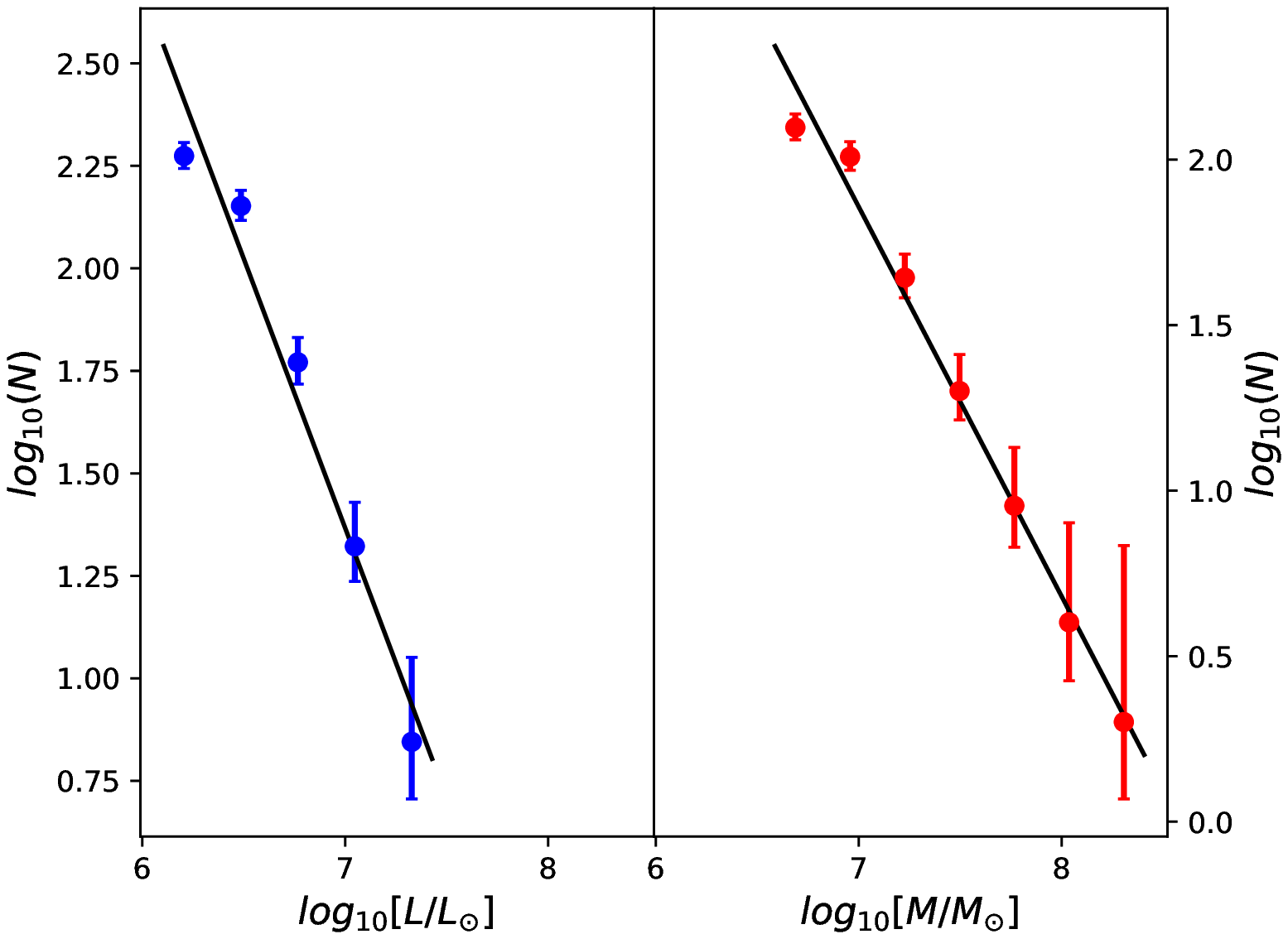}
\caption[Dwarf Members Mass \& Luminosity functions]{Power law fitted to the luminosity function (\textit{left}) and the mass function (\textit{right}). The error bars are based on Poisson errors. The luminosity function has a steeper faint end slope than the Local Group observations with $\sim$5 times the dwarf galaxies observed in the Local Group.}
\label{fig:mass_functions}
\end{figure}

\subsection{Surface Brightness - Absolute Magnitude}\label{sec:sb_Mag}
	We study the surface brightness properties of our dwarf candidates using the accurate shape and size measurements from the S\'ersic profile (Equation \ref{eq:sersic}) fitting. We make use of calculations from \cite{graham2005concise} to convert from intensity to surface brightness profile $\mu(R)$, given by

\begin{equation}\label{eq:muRfull}
\mu(R) = \mu_{e} + \frac{2.5b_n}{ln(10)}\Bigg[\bigg(\frac{R}{R_{e}}\bigg)^{\frac{1}{n}} - 1\Bigg]
\end{equation}
where $\mu_{e} \equiv \mu(R_{e})$ is the surface brightness at the effective radius $R_e$, the S\'ersic index $n$ and the related $b_n$ S\'ersic coefficient. We make use of an analytical approximation of $b_n$ derived in literature, where $b_n = \big(2n-\frac{1}{3}\big) + \mathcal{O}(n^{-1})$ \citep{Ciotti1991}.

	A special case of Equation~\ref{eq:muRfull}, computed for this investigation, also derived in \cite{graham2005concise}, is the `mean effective surface brightness', $\langle\mu\rangle_e$, which is found using the average effective intensity within the radius containing half of the total light
    
\begin{equation}\label{eq:mumean}
\begin{split}
\langle\mu_{e}\rangle &= \mu_e -2.5log[f(n)]
\end{split}
\end{equation}
where $f(n)$, which describes the profile shape, also depends on the complete gamma function, $\Gamma(2n) = 2\gamma(2n,b_n)$.
    
    	Figure~\ref{fig:mueff_mv} shows the surface brightness versus magnitude distribution of our sample of dwarf candidates (black circles). For comparison purposes, we also show the region occupied by the LG galaxies (red stars; \citealt{2012mcconn}). We also show the region occupied by the low surface brightness (LSB) galaxies from the Perseus cluster \citep{wittmann2017perseus} as hollow squares. LSBs are defined as galaxies with $\langle\mu_{e}\rangle_{v} >$ 24.8 mag arcsec\textsuperscript{-2} in R band and the the so-called ultra-diffuse galaxies (UDGs) defined as $\langle\mu_{e}\rangle_{v} \geq$ 24.8 mag arcsec\textsuperscript{-2} and R\textsubscript{e} $\geq$ 1.5kpc \citep{yagi2016catalog} which, have been found to exist in great numbers in clusters \citep{vanDokkum2015UDG,koda2015discovery,yagi2016catalog}.   However, these UDGs are just a subset of dwarf elliptical galaxies found mostly in rich clusters of galaxies (e.g., \citealt{conselice2003perseus,conselice2018udg}).
    
    Implementing the above criteria in Figure \ref{fig:mueff_mv}, we find 53 analogues to the LSB galaxies (green triangles) found in the Perseus cluster core. Recently, a study of abundances of UDGs in lower mass systems, i.e. galaxy groups has been undertaken as part of the Galaxy And Mass Assembly (GAMA) survey \citep{vanderburg2017udg} who find that we should expect $\sim$ 1 in 5 LG mass analogue systems to contain an UDG. Our results are consistent with the expectations from this study. 
    
\subsection{Luminosity Function}\label{sec:lum_func}
    It has been shown that there exists a tight correlation between the stellar mass-to-light ratio (M/L) and the colours of the integrated light from a galaxy, corresponding to its stellar populations \citep{2001Bell}. \cite{2001Bell} evolve various formation models with a Salpeter initial mass function (IMF). Among these models, the mass dependent galaxy formation epoch model, including a burst is found to reproduce the observed local trends in age and metallicity with K-band magnitude from \cite{bell2000stellar}. As such, we use the relation:
\begin{equation}\label{eq:ml_ratio}
\log_{10}\bigg(\frac{M}{L}\bigg) = -0.820 (B - R) + 0.851
\end{equation}
given by this model to compute the stellar masses of the dwarf candidates.

\begin{table*}
\begin{center}
\caption{Truncated dwarf candidate catalogue for the NGC-3175 group. In order: RA, Dec, R band magnitude, B band magnitude, effective radius, S\'ersic index, axis ratio, position angle, goodness of fit and SExtractor area (in pixel$^2$ units).}
\label{tab:cata}
\begin{tabular}{cccccccccc}
\hline
$\alpha(J2000)$ & $\delta(J2000)$ & $m_{R}$ & $m_{B}$ & $R_{e}$ & $n$ & $\frac{b}{a}$ & PA & $\chi^{2}_{\nu}$ & Area \\ 
(deg) & (deg) & ($\pm$ 0.1 mag) & ($\pm$ 0.1 mag) & ($\pm 27.2 $pc) & {$\pm$ 0.1} & {} & (deg) & {} & (pix$^2$) \\ 
\hline \vspace{0.1cm}
153.886865 & -29.210021 & 19.9 & 21.4 & 162.9 & 0.9 & 0.13 & 20 & 1.2 & 80 \\
153.580018 & -29.206598 & 18.0 & 19.2 & 239.6 & 0.9 & 0.25 & 50 & 1.1 & 745 \\
153.727596 & -29.189941 & 19.9 & 21.5 & 161.4 & 1.5 & 0.32 & 59 & 1.1 & 144 \\
153.426678 & -29.052451 & 20.4 & 22.3 & 132.0 & 0.4 & 0.44 & 69 & 1.1 & 101 \\
153.530832 & -29.042335 & 20.4 & 21.3 & 152.6 & 0.3 & 0.15 & 23 & 1.2 & 110 \\
{...} & {} & {} & {} & {} & {} & {} & {} & {} & {} \\
153.62877 & -29.177252 & 19.1 & 20.0 & 147.8 & 0.8 & 0.28 & -22 & 1.1 & 358 \\
153.439392 & -29.163128 & 20.3 & 21.4 & 131.1 & 1.1 & 0.54 & 75 & 1.0 & 69 \\
153.431708 & -29.161917 & 20.2 & 21.0 & 145.0 & 0.9 & 0.44 & 9 & 1.0 & 84 \\
153.309428 & -29.14232 & 20.2 & 21.8 & 124.8 & 1.1 & 0.55 & 4 & 1.2 & 95 \\
153.837497 & -29.100491 & 19.4 & 20.3 & 142.3 & 0.5 & 0.47 & -82 & 1.3 & 315 \\
\hline
\end{tabular}
\end{center}
\end{table*}

\begin{table}
\begin{center}
\caption{The parameters from the linear fit of the form in Equation~\ref{eq:plaw_fit} to the mass and luminosity function. The 1$\sigma$ uncertainties on the fitted parameters are also shown.}
\label{tab:functioncoeff}
 \begin{tabular}{lcr}
 \hline
 Function & $\alpha$ & C \\ [0.5ex] 
 \hline
 Luminosity & {-1.31 $\pm$ 0.25} & {10.57 $\pm$ 1.69} \\ 
 Mass & {-1.18 $\pm$ 0.07} & {10.09 $\pm$ 0.54} \\ [1ex] 
 \hline
\end{tabular}
\end{center}
\end{table}

	For our data, we fit a linear function of the form
\begin{equation}
\log(N) = \alpha \log(L) + C
\label{eq:plaw_fit}
\end{equation}
    where $\alpha$ is now the slope describing the shape of the power law and $C$ is a constant. We fit a function of this form to both the luminosity and the mass function as shown in Figure \ref{fig:mass_functions} (\textit{left}) and (\textit{right}) respectively. We do not plot any points after incompleteness limit is reached. The error bars are calculated using Poisson statistics. The summary of the slope parameters calculated by this linear fit are shown in Table \ref{tab:functioncoeff}. Both the luminosity and mass functions are not normalised with the volume, however this does not affect the slope parameter $\alpha$, which can be used for comparison with previous studies. We find the faint-end slope $\alpha$ of the LF to be steeper than that observed in the Local Group.
    
\section{Cosmological Implications}\label{sec:6}
	We now present the results of our investigation in the context of the missing satellite problem (MSP) (e.g., \citealt{moore1999dark}). Due to the similarities with the LG, we can compare the properties and the LF of the observed dwarf candidates of the NGC-3175 group with those predicted from numerical simulations of the Local Group carried out by others easily. We study satellite abundances as a function of stellar masses.

\begin{figure*}
\centering
\includegraphics[width=\textwidth]{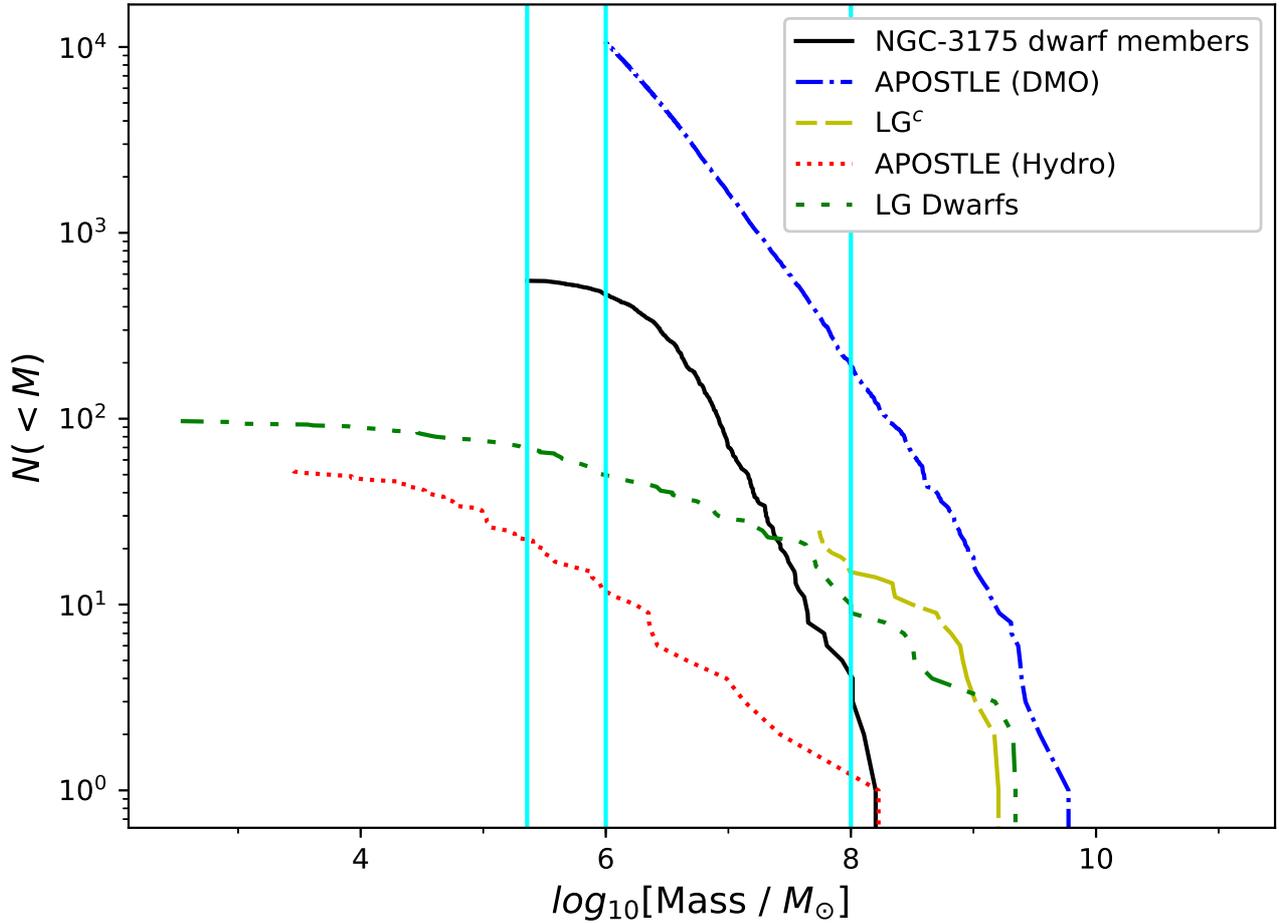}
\caption[Cumulative number of satellites]{The cumulative number count of the dwarf members of the NGC-3175 group are presented against stellar masses. To contrast the observed catalogue, literature data for LG dwarfs (red dotted line) from \cite{2012mcconn} are shown. In addition, the constrained MW+M31 hydrodynamical simulation (yellow line) from \cite{scannapieco2015constr} and, the APOSTLE AP04H (DMO) dataset (blue line) and the APOSTLE AP04H (red line) hydrodynamical run \citep{fattahi2016apostle} are also shown. The simulation data is shown only for haloes/galaxies within the central $\sim$500kpc of the centre of the group. The vertical cyan lines are drawn to aid in comparison of our data with the over-plot literature data. We observe $\sim$5 times the dwarf candidates as found in the Local Group at the faint end.}
\label{fig:Cumu_plots}
\end{figure*}

	In Figure~\ref{fig:Cumu_plots}, we plot the cumulative satellite abundances for the dwarf candidates of the NGC-3175 group (black line). In addition, we also plot the abundances of the observed dwarf galaxies around the MW and M31 (green dot-dashed line) using data complied by \citep{2012mcconn}. For simulation data, we overlay a dark matter only (DMO) dataset (blue dashed line; hereafter AP04) which, comes from the APOSTLE (level AP04) Local Group simulation \citep{fattahi2016apostle} with a resolution of $10^{6} M_{\odot}$. We also show data from the hydrodynamical run of this project (red dotted line; hereafter AP04H). In both cases, we restrict the data to haloes/galaxies within a sphere of radius 500kpc from the centre of the group. In yellow, we plot the output from a constrained simulation of the formation of MW+M31 (hereafter LG\textsuperscript{C}; \citealt{scannapieco2015constr,creasey2015constr}). Also plotted in cyan are vertical lines to aid with comparisons to our data.
    
    Although the observed satellite abundances in the LG roughly match the output from the AP04H simulation (when not restricted to haloes within 500kpc), the DMO luminosity function is much steeper than the observed. This is expected as not all of the low mass dark matter haloes will host a galaxy due to feedback effects, for example, from supernovae. The number of dwarf candidates found at the higher end of masses ($ > 10^{8} M_{\odot}$) in our sample is lower than that of the observed LG population which suggests that we may be missing some very bright/massive galaxies from our sample, as suggested in Section~\ref{sec:M_vs_re}. Interestingly, the same is true for all but the AP04H run, which is less steep than our observations at this mass scale. At the depth of our imaging data (low mass end), we detect $\sim$ 20 times more dwarf candidates than galaxies predicted by AP04H run and a factor of $\sim$ 5 times more than the LG observations.
    
	It is likely that some of our dwarfs are due to contamination and are not true satellite galaxies, but it is unlikely that enough are due to contamination to match the Local Group abundances given our group membership criteria in Section~\ref{sec:bkg_gals}. It tentatively appears that at the limit of our data, our abundance observations lie in between a hydrodynamical and a dark matter only simulation. To resolve the issue of group membership, we are obtaining spectrospcopy for some of the dwarf candidates identified in this study.

\section{Conclusions}\label{sec:7}
	We identify and study the properties of the dwarf satellite members of the NGC-3175 group (a Local Group analogue) within its central $\sim$ 500kpc down to M$_B \sim$ -7.7. The catalogue of dwarf candidates of NGC-3175 group is used to investigate the missing satellite problem by comparing the observed luminosity function with Local Group observations and predictions from a range of $\Lambda$CDM simulations.
    
    We use SExtractor to detect dwarf galaxies in our R and B-band imaging data (Section~\ref{sec:3}), fit surface brightness profiles to determine accurate structural properties using GALFIT and, determine the subset of likely dwarf candidates of the group based on the total colour, surface brightness and morphological visual inspection (Section~\ref{sec:4}). Through this analysis, we identify 553 dwarf candidates of the group, most of which have never been studied before.
    
    Perhaps unsurprisingly, we find that the dwarf galaxies appear to have similar properties (lower mass, fainter and smaller) to the Low Mass Cluster Galaxies (LMCGs) found near the central regions of clusters (e.g. Perseus: \citealt{conselice2002galaxy}). The NGC-3175 dwarf galaxies occupy a small region in the size-absolute magnitude plane, showing a change in size by an order of magnitude for a magnitude change of $\sim$5mag. From this relation, we also find that on average, dwarf candidates from this study are smaller than the dwarf satellites of the MW and M31. This may indicate a possible selection bias that preferentially selects smaller sized dwarf galaxies, or we have measurements which are more accurate resulting in a lower scatter. However, performing the same analysis on simulated images shows no such bias in the output of model fitting. This indicates that the observed trend of smaller sizes dwarf galaxies is likely not due to our analysis but instead has an astrophysical origin.
    
    We study the implications to the standard cosmological model by characterising the luminosity function of the dwarf candidates down to M\textsubscript{B} $\sim$ -7.7 and compare with observations and simulations of the Local Group. We find that at the bright end (M\textsubscript{R} < -12), where there exists a census for Local Group dwarfs, the abundances of our dwarf galaxy population roughly matches the abundances of Local Group observations and predictions. However, at the (faint) limit of our data, we identify more than 5 times the observed Local Group satellite population. In comparison with predictions from simulations, our observed satellite abundances lie between a dark matter only model and a hydrodynamical $\Lambda$CDM model. However, that we detect far more dwarf galaxies than both observations and simulations may be expected if the mass density of our group is different to that of the Local Group. This can also have an effect on the galaxy sizes of the satellite population, for example, the satellites of M31 are consistently smaller than that of the Milky Way \citep{2012mcconn}.
    
    Observations of the Local Group have already found satellite galaxies more than 5 mag fainter than is possible with our imaging data around the NGC-3175 group, making the problem even more severe. Our results suggest that the Local Group satellite population may not be representative of other groups in the wider Universe. Some of the discrepancy between the observations and simulations may be from contamination due to background objects in our sample. However, it is unlikely that there are enough candidates due to contamination to match the observed Local Group population after performing our morphological, colour and surface brightness based selection of group members. 
    
    We are now obtaining spectroscopy for some of these dwarf candidates to determine how many are real dwarf members as opposed to background galaxies which will settle this issue. The spectra will also allow us to measure the mass density of this group to allow a fairer comparison to the Local Group. Recently, observations have also shown that satellite galaxies lie along a plane in the Local Group around both the Milky Way \citep{2012MNRAS.423.1109P,2013MNRAS.435.2116P} and M31 \citep{2013ApJ...766..120C,2013Natur.493...62I}, contrary to the expectation of an isotropic distribution of satellites from our understanding of hierarchical structure formation \citep{2017MNRAS.466.3119A}. Evidence for the existence of a plane of satellites around a system beyond the Local Group has also been found (Centaurus A: \citealt{2018Sci...359..534M}). However, recent results from Gaia DR2 cast the existence of a plane of satellites into doubt for the Local Group \citep{2018arXiv180500908F}, and thus further work on this topic is needed. Velocity measurements from spectra will also allow us to test this deviation from $\Lambda$CDM for the NGC-3175 group.


\section*{Acknowledgements}
The authors would like to acknowledge support from the Science and Technologies Facilities Council (STFC) that made this study possible. This study contains data based on observations collected at the European Organisation for Astronomical Research in the Southern Hemisphere under ESO programme(s) 086.A-0589(A). The authors would like to extend their gratitude to Frazer Pearce and Steven Bamford for their expertise when discussing simulations and GALAPAGOS-2, respectively. The authors also thank Philip Best for providing feedback and suggestions which improved the paper. We would also like to thank Cecilia Scannapieco and their respective colleagues for making their simulation data available for comparison. We also thank the undergraduates O. Marshall and S. De Silva who processed the raw images. SJP was supported by an STFC studentship at the time of the telescope proposal submission. SJP also acknowledges support from ESO for the observing trip at La Silla. We also thank the anonymous referee for their comments and suggestions, which clarified and improved the presentation of this paper.

We thank the APOSTLE and EAGLE teams for access to unpublished data from the APOSTLE simulations of the Local Group, which were carried out with the EAGLE code. The APOSTLE simulations were supported by the Science and Technology Facilities Council [ST/J501013/1, ST/L00075X/1]. They used the DiRAC Data Centric system at Durham University, operated by the Institute for Computational Cosmology on behalf of the STFC DiRAC HPC Facility www.dirac.ac.uk). This equipment was funded by BIS National E-infrastructure capital grant ST/K00042X/1, STFC capital grant ST/H008519/1, and STFC DiRAC Operations grant ST/K003267/1 and Durham University. DiRAC is part of the National E-Infrastructure.




\bibliographystyle{mnras}
\bibliography{references} 







\label{lastpage}
\bsp	

\end{document}